\documentclass[aps,prl,twocolumn,superscriptaddress,floatfix]{revtex4-2}

\usepackage[utf8]{inputenc}
\usepackage{amsmath,amssymb}
\usepackage{physics}
\usepackage{esvect}
\usepackage{bm}

\usepackage{graphicx}
\usepackage{subfig}   % <-- APS-compatible subfigure package
\usepackage{tikz}
\usepackage[percent]{overpic}

\usepackage{ragged2e}
\usepackage{microtype}

\usepackage{cleveref}

\begin{document}

\title{Less is more: subspace reduction for counterdiabatic driving of Rydberg atom arrays}

\author{Wen Ting Hsieh}
\affiliation{Department of Physics, New York University, New York, NY, USA}

\author{Dries Sels}
\affiliation{Department of Physics, New York University, New York, NY, USA}
\affiliation{Center for Computational Quantum Physics, Flatiron Institute, New York, NY, USA}

\date{\today}

\begin{abstract}
This study explores the use of subspace methods in combination with counterdiabatic driving in a Rydberg atom system to solve the Maximum Independent Set (MIS) problem. Although exact counterdiabatic driving offers excellent performance, it comes at an unscalable computational cost. In this work, we demonstrate that counterdiabatic driving can be significantly improved by restricting the analysis to a relevant subspace of the system. We first show that both direct diagonalization and the Krylov method for obtaining the counterdiabatic matrix can be accelerated through the use of subspace techniques, while still maintaining strong performance. We then demonstrate that the cost function used in the standard Krylov method can be further optimized by employing a subspace-based cost function. These findings open up new possibilities for applying counterdiabatic driving in a practical and efficient manner to a variety of quantum systems.
\end{abstract}

\maketitle

\section{Introduction}
\label{Intro}

Adiabatic Quantum Computation (AQC) is a quantum computing paradigm designed to solve complex computational problems. It gradually evolves the Hamiltonian of a quantum system from an initial state, characterized by an easily prepared ground state, to a final Hamiltonian whose ground state encodes the solution to the targeted problem. Provided that this Hamiltonian evolution proceeds sufficiently slowly, the quantum system remains close to its instantaneous ground state throughout the entire process, enabling AQC to efficiently address challenging computational tasks~\cite{kadowaki1998quantum, farhi2000quantum, farhi2001quantum, aharonov2008adiabatic, johnson2011quantum, lucas2014ising,  vinci2016nested, albash2018adiabatic, weinberg2020scaling, king2022coherent, trummer2015multiple, king2025beyond, orquin2025analog}. AQC exhibits potential quantum speedups for certain problem classes and can simulate any circuit-model algorithm with at most polynomial overhead. Additionally, its inherent robustness against decoherence can facilitate transitions to the ground state, particularly in low-temperature environments~\cite{mcgeoch2014adiabatic}. However, the efficiency of AQC depends on the minimum energy gap, with annealing time scaling inversely with its square~\cite{farhi2000quantum}, or inversely with the gap under an optimized adiabatic protocol when one has detailed knowledge of the ground state geometry, which remains optimal when restricted to the original set of control Hamiltonians~\cite{bukov2019geometric}. Therefore, when the energy gap diminishes exponentially, as is often the case in large-scale quantum systems, the computational time increases significantly, limiting both the practical and fundamental applicability of these methods~\cite{hogg2003adiabatic, altshuler2010anderson}. It is thus crucial to develop approaches that address this limitation.

Transitionless driving~\cite{berry2009transitionless, demirplak2003adiabatic, demirplak2005assisted, demirplak2008consistency, zheng2016cost}, often referred to as counterdiabatic driving, provides an effective approach to addressing the challenges associated with adiabatic quantum computation, particularly the issue of prolonged computational times due to diminishing energy gaps. This technique offers several significant advantages and applications~\cite{pandey2020adiabatic,  guery2019shortcuts, del2012assisted, jarzynski2013generating, del2013shortcuts, tseng2013counterdiabatic, takahashi2013transitionless, kolodrubetz2017geometry, hartmann2019rapid, chung2019shortcuts, hartmann2020multi, nakahara2022counterdiabatic,  schindler2024counterdiabatic, vcepaite2024counterdiabatic, gangopadhay2025counterdiabatic, huerta2025quantum, vithanage2025fast}, notably its ability to prevent unwanted transitions between instantaneous eigenstates, thus reducing losses during state evolution~\cite{TORRONTEGUI2013117, hegade2021shortcuts, hegade2022digitized, chandarana2022digitized, yague2023shortcut, vcepaite2023counterdiabatic, duncan2024exact, morawetz2024efficient, takahashi2024shortcuts, lukin2024quantum, simen2025digitized, romero2025bias, hatomura2024shortcuts}. Moreover, by employing exact counterdiabatic driving, arbitrarily fast annealing protocols can be realized, enabling rapid modifications of the system Hamiltonian and thus avoiding coupling to the environment to spoil the quantum evolution. Recent advances further demonstrate that counterdiabatic techniques can circumvent topological defects, thereby improving computational robustness and performance~\cite{visuri2025digitized}, and can also enhance sampling efficiency~\cite{hegade2025digitized}.

Despite the significant advantages offered by counterdiabatic driving, it is widely acknowledged that determining the exact counterdiabatic terms is computationally challenging and can sometimes be more difficult than solving the original problem itself. The primary difficulty arises from the complexity of diagonalizing the Hamiltonian, a process that requires computational resources scaling exponentially with system size. This challenge has motivated numerous studies aimed at developing methods to approximate counterdiabatic terms in a computationally efficient manner~\cite{saberi14, kim2024variational, zhang2024analog, bottarelli2024symmetry, passarelli2020counterdiabatic, van2024gate}, as well as designing strategies to facilitate their implementation in practical settings ~\cite{prielinger2021two, shende2024experimental, vizzuso2024role, li2024quantum}. Recently, a universal and efficient scheme was proposed for systems with a finite gap~\cite{morawetz2025universal, finzgar25}.

One of the approaches to simplifying the search for the counterdiabatic term uses a Krylov expansion of the instantaneous counterdiabatic matrix, which is then approximated using a limited number of nested commutators~\cite{sels2017minimizing, hegade2022digitized, hegade2025digitized}, forming a Krylov subspace of operators~\cite{liesen2013krylov, nandy2025quantum}. While the exact counterdiabatic term may involve a set of operators whose number increases exponentially and may be difficult to implement experimentally, the Krylov form obtained from the variational method presents several advantages~\cite{claeys2019floquet, sels2017minimizing}. From an experimental perspective, the Krylov method provides an efficient approximation of the exact gauge potential using a limited set of parameters and offers a practical route for implementation via a Floquet engineered approach.

In this paper, we introduce a method that significantly improves the efficiency of finding the counterdiabatic term, building on the scheme of the Maximum Independent Set (MIS) problem. We explore the potential to simplify the counterdiabatic driving by employing a subspace-based formulation. This approach allows for a more efficient determination of both the exact counterdiabatic term and the Krylov approximation, while naturally reducing the computational overhead. Within this framework, we also propose an alternative strategy to improve the fidelity of the final state through the use of submatrix elements as the cost function in the variational procedure. We show that, by restricting the cost function to a relevant subspace rather than the full Hilbert space, one can achieve improved final fidelity with reduced computational complexity.

\section{Methodology}
\label{sec:Methodology}
\subsection{Counterdiabatic Driving}\label{sec:CD}
The time evolution of a closed quantum system is governed by the time-dependent Schr\"{o}dinger equation:
\begin{equation}
    i\hbar\dv{}{t}\ket{\psi(t)}\;=\;H_{0}(t)\ket{\psi(t)}
\end{equation}
$H_{0}(t)$ is, in general, a time-dependent Hamiltonian. In the context of adiabatic computing, the ground state of $H_{0}(t_0)$ at the initial time $t_0$ is assumed to be easy to prepare, while the ground state of $H_{0}(t_f)$ at the final time $t_f$ encodes the solution to the problem of interest. To better analyze the system's dynamics, we perform a transformation to the instantaneous eigenbasis of the Hamiltonian. This is achieved through the following unitary transformation:
\begin{equation}
    \tilde{\ket{\psi(t)}}=U^{\dag}(t)\ket{\psi(t)}
\end{equation}
and
\begin{equation}
    \tilde{H_{0}}(t)=U^{\dag}(t)H_{0}(t)U(t)
\end{equation}
\noindent where $U(t)$ is a unitary matrix whose columns are the instantaneous eigenvectors of the Hamiltonian $H_{0}(t)$, and $U^{\dag}(t)$ denotes its Hermitian conjugate. The tilde notation (e.g., $\tilde{\ket{\psi(t)}}$) indicates that the corresponding quantities are represented in the instantaneous eigenbasis. The time evolution of the wavefunction in this basis is then given by:
\begin{align}\label{eq:schr_eigen}
    i\hbar\dv{}{t}\tilde{\ket{\psi(t)}} 
    =\;\tilde{H}(t)\tilde{\ket{\psi(t)}}
\end{align}
where the effective Hamiltonian $\tilde{H}(t)$ in this time-dependent basis can be expressed as:
\begin{align}\label{eq:H_eigen}
    \tilde{H}(t) &= \tilde{H_{0}}(t) + i\hbar\dot{U^{\dag}}(t)U(t) \nonumber \\
    &= \tilde{H_{0}}(t) + i\hbar\dot{\lambda}\frac{\partial U^{\dag}(t)} {\partial \lambda}U(t) \nonumber \\ 
    &= \tilde{H_{0}}(t)-\dot{\lambda}(t)\tilde{A_{\lambda}}(t)
\end{align}
This is the direct result of a unitary transformation of the original Schr\"{o}dinger equation. The purpose of this transformation is to separate the interstate transitions. Since $\tilde{H_{0}}(t)$ is diagonal in its own eigenbasis, all non-adiabatic transitions between eigenstates arise from the gauge potential $\tilde{A_{\lambda}}(t)$~\cite{kolodrubetz2017geometry}:
\begin{equation}\label{eq:gauge}
    \tilde{A_{\lambda}}(t) = -i\hbar\frac{\partial U^{\dag}(t)}{\partial \lambda}U(t)
\end{equation}
which in the lab frame reads
\begin{equation}
    A_{\lambda}(t)=U(t)\tilde{A_{\lambda}}(t)U^{\dag}(t)=-i\hbar U(t)\frac{\partial U^{\dag}(t)}{\partial \lambda}
\end{equation}
To suppress these non-adiabatic interstate transitions, a counterdiabatic driving term is introduced, modifying the system by replacing the original Hamiltonian $H_{0}(t)$ with: 
\begin{equation}\label{eq:CD_eq0}
H_{CD}(t)=H_{0}(t)+\dot{\lambda}(t)A_{\lambda}(t)
\end{equation}
This additional term cancels the off-diagonal components in Eq.~\eqref{eq:H_eigen}, resulting in a purely diagonal Hamiltonian in the eigenbasis, $\tilde{H}(t)$=$\tilde{H_{0}}(t)$. Therefore, a system initialized in an $n$th eigenstate of the initial Hamiltonian, $\ket{\psi(0)}=\ket{E_{n}(0)}$, where the eigenstates are ordered by increasing energy, as is conventional, evolves under the counterdiabatic driving Hamiltonian into the corresponding $n$th instantaneous eigenstate at time $t$:
\begin{equation}
\ket{\psi(t)}=e^{i\zeta_{n}(t)}\ket{E_{n}(t)}
\end{equation}
\noindent where $\zeta_{n}$ is the adiabatic phase accumulated by the 
$n$th energy eigenstate during the evolution~\cite{berry2009transitionless}.

\subsection{Variational Principle and Krylov Method}
From the derivation of gauge potential $A_{\lambda}$ in Eq.~\eqref{eq:gauge}, it is evident that obtaining the exact form of the gauge potential requires direct diagonalization. To alleviate the computational complexity of this process and limit the number of operators required to construct the exact gauge potential~\cite{claeys2019floquet,takahashi2024shortcuts}, an approximation method known as Krylov expansion is employed, as shown below:
\begin{equation}\label{eq:A}
A_{\lambda}^{(l)}=i\sum_{k=1}^{l}\alpha_{k}O_{\lambda, k}
\end{equation}
\noindent where each $O_{\lambda, k}$ represents an operator of the form: 
\begin{equation}
O_{\lambda, k}=\underbrace{[H_{0},[H_{0},...[H_{0},}_\text{2k-1}\pdv
{H_{0}}{\lambda}]...]]
\end{equation}
\noindent with a total of $2k-1$ nested commutators. To determine the optimal coefficients ${\alpha_{k}}$, we adopt a variational principle as introduced in Ref.~\cite{sels2017minimizing}. Specifically, we defined the action as:
\begin{equation}\label{eq:Sl}
S_{l}(A^{(l)}_{\lambda})={\rm Tr} \left[G^{2}_{\lambda}(A_{\lambda}^{(l)})\right]
\end{equation}
Minimizing this action is equivalent to minimizing the Frobenius norm of the operator $G_{\lambda}$, defined as:
\begin{equation}\label{eq:G}
G_{\lambda}(A_{\lambda}^{(l)})=\pdv{H_{0}}{\lambda}+\frac{1}{i\hbar}[H_{0},A_{\lambda}^{(l)}]
\end{equation}
The action $S_{l}$ is minimized with respect to the coefficients $\alpha_{k}$ to ensure the best approximation to the exact gauge potential. This leads to the set of variational conditions:
\begin{equation}\label{eq:dSl_da}
\pdv{S_{l}(A^{(l)}_{\lambda})}{\alpha_{i}}=0
\end{equation}
\noindent which results in the following equation:
\begin{align}\label{eq:mincost}
& \left\langle \pdv{H_0}{\lambda}, Q_{\lambda, i} \right\rangle_F
+ \left\langle Q_{\lambda, i}, \pdv{H_0}{\lambda} \right\rangle_F \notag \\
& + \sum_{j=1}^{l} \alpha_j \left(
\left\langle Q_{\lambda, i}, Q_{\lambda, j} \right\rangle_F
+ \left\langle Q_{\lambda, j}, Q_{\lambda, i} \right\rangle_F \right)
= 0
\end{align}
\noindent with $\left<A,B\right>_{F}={\rm Tr}(A^{\dag}B)$ denoting the Frobenius inner product. We define $Q_{\lambda, k}$ as an operator with an even number of commutators:
\begin{equation}\label{eq:Q}
Q_{\lambda, k}=[H_{0}, O_{\lambda, k}]=\underbrace{[H_{0},[H_{0},...[H_{0},}_\text{2k}\pdv
{H_{0}}{\lambda}]...]]
\end{equation}
\noindent This procedure yields a variationally optimal approximation to the adiabatic gauge potential within the space spanned by the Krylov basis ${O_{\lambda, k}}$.
\begin{figure}[t]
    %\centering
    % -------- (a) Top figure --------
    \begin{minipage}[b]{0.9\linewidth}
        \begin{picture}(0,0)
            \put(-127,-10){\textbf{(a)}}  % move label: left/right & up/down
        \end{picture}
        \includegraphics[width=\linewidth]{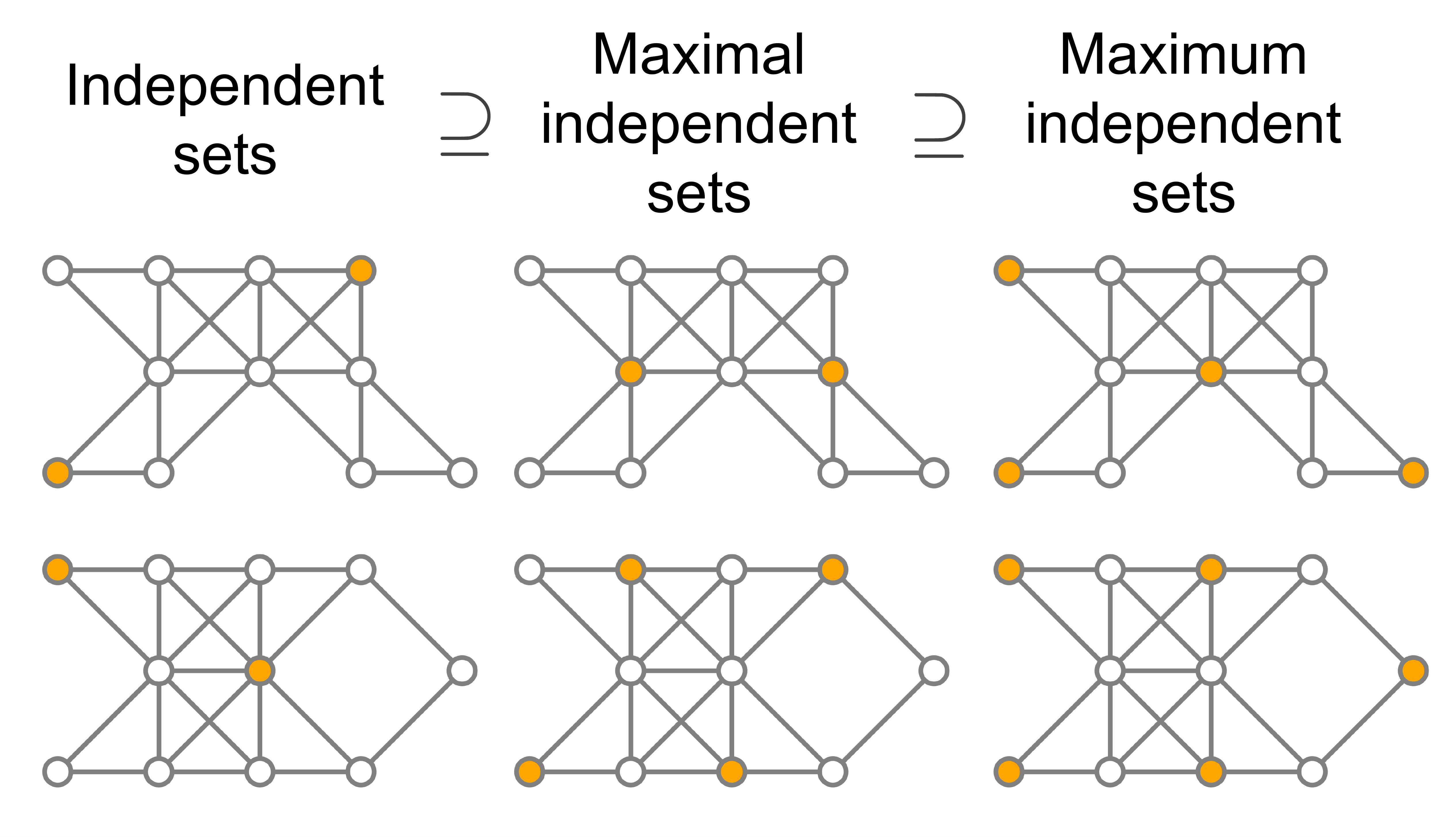}
    \end{minipage}
    \vspace{0.25cm}  

    % -------- Bottom two --------
    \begin{minipage}[b]{0.75\linewidth}
        \begin{picture}(0,-0)
            \put(-127,-10){\textbf{(b)}}  % adjust -10,10 if needed
        \end{picture}
        \includegraphics[width=\linewidth]{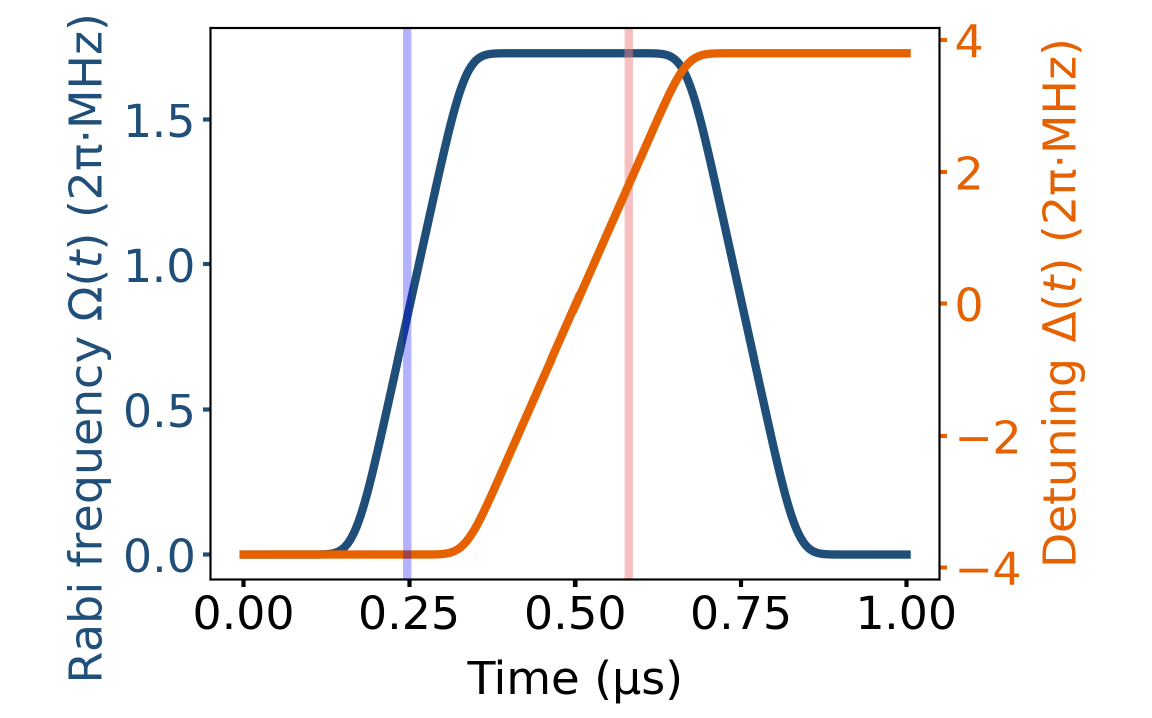}
    \end{minipage}
    \vspace{-0.4em}
    \caption{\protect\justifying (a) Representative 11-vertex graphs illustrating independent sets, maximal independent sets, and maximum independent sets, with each example highlighting the distinctions among these categories. (b) Example control waveforms for the Rabi frequency $\Omega(t)$ and detuning $\Delta(t)$ used in maximum independent set searches. Vertical blue and light-coral lines indicate the time slices used for evaluating the corresponding adiabatic gauge-potential matrices discussed later.}
    \label{fig:IS}
\end{figure}
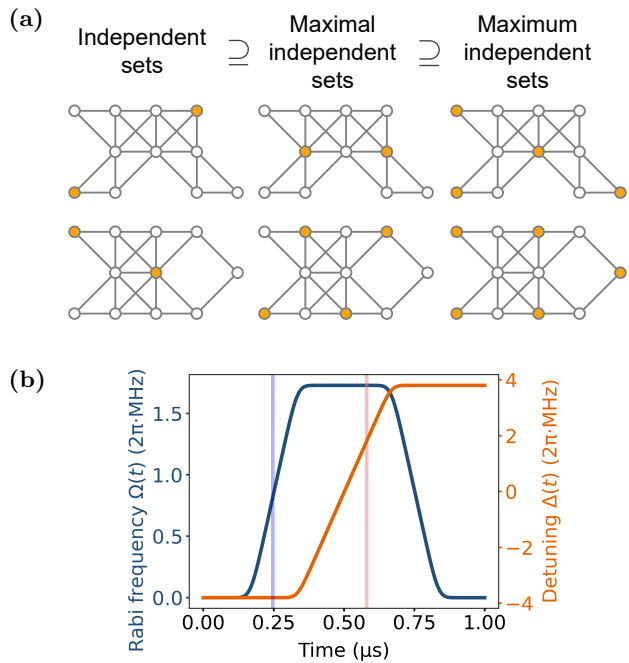

\subsection{Maximum Independent Set}
Having introduced the basic concepts, we focus on solving the maximum independent set (MIS) problem. An independent set in a graph is a set of vertices such that no two vertices are adjacent. A maximal independent set is an independent set that cannot be extended by adding any additional vertex without violating the independence condition. A maximum independent set is a maximal independent set with the largest possible number of vertices, and identifying it is the goal of this study. An illustration of independent, maximal independent, and maximum independent sets is provided in Fig.~\ref{fig:IS}(a). The independent set decision problem is polynomially equivalent to the Clique decision problem via graph complementation. Since Clique is NP-complete~\cite{karp2009reducibility, childs2000finding}, independent set is also NP-complete. Consequently, the optimization version, which asks for a maximum independent set, is NP-hard. Moreover, the maximum independent set problem is also hard to approximate, and this hardness persists even for sparse or bounded-degree graphs.

Neutral atom platforms are widely regarded as a natural setting for simulating the MIS problem~\cite{ebadi2022quantum, pichler2018quantum, nguyen2023quantum, nguyen2023quantum, sohrabizadeh2024gnn, bombieri2025quantum, schuetz2025qredumis, manovitz2025quantum, ebadi2021quantum, kim2022rydberg, sohrabizadeh2024gnn, beterov2025counterdiabatic}. We simulated the time evolution of neutral-atom quantum systems to address this problem. Specifically, the Hamiltonian corresponding to a given graph, similar to  Fig.~\ref{fig:IS}(a), is designed to encode the MIS solution:
\begin{equation} \label{eq:H_ryd}
H_{0}(t) = \sum_{i>j}{\frac{C_{0}}{|r_{i}-r_{j}|^{6}}n_{i}n_{j}} + \Omega(t)\sum_{i}\sigma_{x} - \Delta(t)\sum_{i}n_{i}
\end{equation}
\noindent The Hamiltonian consists of three terms, each playing a distinct role in the dynamics. The first term describes the van der Waals interaction between atoms in the Rydberg state, commonly referred to as the Rydberg blockade. Here, $C_{0}=2\pi\times8.6269\times10^{5} MHz\; \mu m^{6}$ is the Rydberg interaction coefficient for Rubidium-87 with principal quantum number $n = 70$, orbital angular momentum $S$ ($\ell = 0$), and total angular momentum $J = \tfrac{1}{2}$. Because the interaction strength grows rapidly with the principal quantum number, it can become large enough to prevent adjacent atoms from being excited simultaneously, thereby enforcing the blockade constraint. The operator $n=\ket{1}\bra{1}$ denotes the number operator for the Rydberg state $\ket{1}$. The second term, proportional to the time-dependent Rabi frequency $\Omega(t)$, drives coherent transitions between the ground state $\ket{0}$ and the Rydberg state $\ket{1}$. The third term, governed by the detuning $\Delta(t)$, which arises from the offset of the laser frequency from resonance, modulates the energy landscape of the system to guide it toward the maximum number of Rydberg excitations. The waveforms corresponding to $\Omega(t)$ and $\Delta(t)$ are shown in Fig.~\ref{fig:IS}(b), are designed to evolve the system adiabatically from the state with no Rydberg excitations to the MIS solution~\cite{ebadi2022quantum}.

Even though one can naturally generate only unit-disk graphs, the MIS problem on such graphs remains NP-hard~\cite{andrist2023hardness}. For simulations, we used the Bloqade.jl package created by QuEra Computing Inc.~\cite{bloqade2023quera}, which provides efficient tools for quantum simulation. In this work, we focused on the king’s lattice graph, which includes edge crossings and presents a higher level of computational difficulty. This structure contrasts with the square lattice, which lacks edge crossings and is therefore a planar and non-universal graph.

\section{Results}
\label{Results}
For the Rydberg blockade in solving the MIS problem, we use counterdiabatic method introduced earlier in Eq.~\eqref{eq:CD_eq0} combined with the Rydberg blockade equation Eq.~\eqref{eq:H_ryd}.
The counterdiabatic driving protocol is defined as:
\begin{equation}\label{eq:HCD_ryd}
H_{CD}(t)=H_{0}(t)+\dot{\Omega}(t)A_{\Omega}(t)+\dot{\Delta}(t)A_{\Delta}(t)
\end{equation}
We will simply refer to the second and third term, $\dot{\Omega}(t)A_{\Omega}(t)$ and $\dot{\Delta}(t)A_{\Delta}(t)$, as the Rabi-drive counterdiabatic term and the detuning counterdiabatic term.

\subsection{Efficient Gauge Potential Finding}
\subsubsection{Exact counterdiabatic method}
Here, we attempt to compute the counterdiabatic term in a significantly more manageable way by solving the Hamiltonian restricted to an indepedent set subspace that excludes configurations with simultaneously excited nearest-neighbor and next-nearest-neighbor atoms: 
\begin{equation}
H_{0, IS} = P_{IS}H_{0}P_{IS}
\end{equation}
where $P_{IS}$ is a projector to an independent set subspace
\begin{equation}\label{eq:PIS}
P_{IS}=U_{IS}U_{IS}^T
\end{equation}
\begin{equation}
U_{IS}=
\begin{bmatrix}
| & | & & | \\
u_{IS, 1} & u_{IS, 2} & \cdots & u_{IS, d_{IS}} \\
| & | & & |
\end{bmatrix}
\qquad
\end{equation}
and $\{u_{IS, i}\}$ are simply standard basis vectors of the independent set subspace, for example $[1, 0, 0, 0...]^T$ or $[0, 1, 0, 0...]^T$ if the first and second configuration states satisfy the independent set requirement. $d_{IS}$ is the dimension of independent set subspace, which is the number of configuration states that fulfill the independent set requirement. 

We analyze the effectiveness of this protocol by studying the final fidelity:
\begin{equation}
F_{s}(T)=\sum_{i=1}^{d_{MIS}}\abs{\bra{\psi(T)}\ket{\psi_{MIS, i}
}}^{2}
\end{equation}
where the $\ket{\psi(T)}$ is the final wavefunction obtained by time-evolving the initial ground state $\ket{\psi(0)}=[1, 0, 0, 0...]^T$, which corresponds to no Rydberg excitation. The $\ket{\psi_{MIS, i}}$ are MIS solution states for each specific problem, $d_{MIS}$ is the MIS degeneracy that corresponds to the lowest few energy states. It is worth noting that although this situation very rarely occurs, special care is taken to include only those MIS configurations whose energies are lower than any non-MIS configurations. This condition is stricter than summing over all applicable MIS states, yet it aligns with the purpose of counterdiabatic driving. Therefore, we adopt this restriction to ensure an effective fidelity evaluation of counterdiabatic driving.

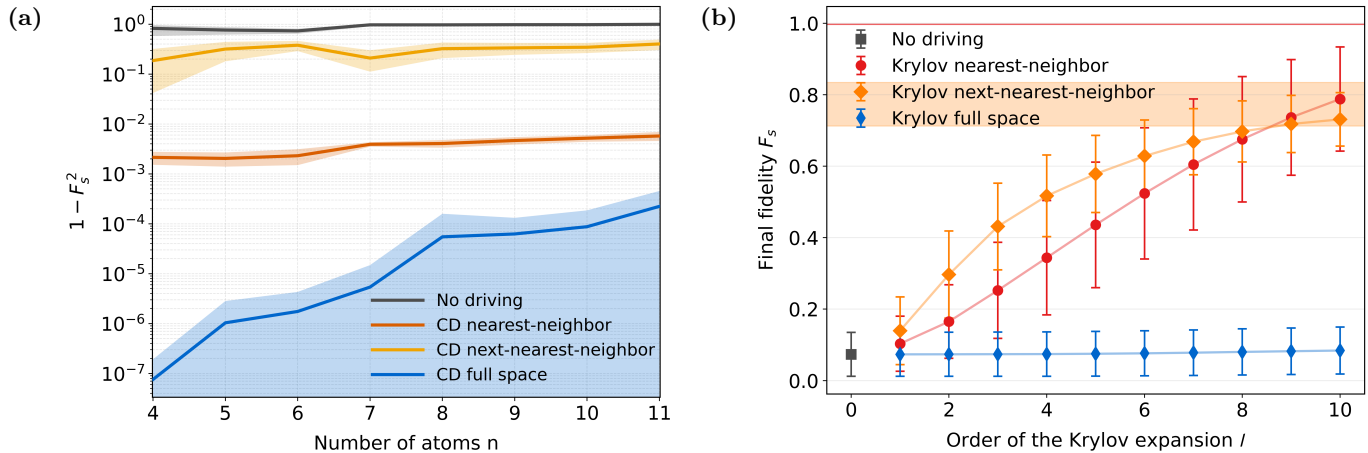
\begin{figure*}[htbp]
    \centering
    \begin{minipage}[t]{0.48\textwidth}
        \begin{picture}(0,0)
            \put(-136,-15){\textbf{(a)}}  % move label: left/right & up/down
        \end{picture}
        \includegraphics[width=\linewidth]{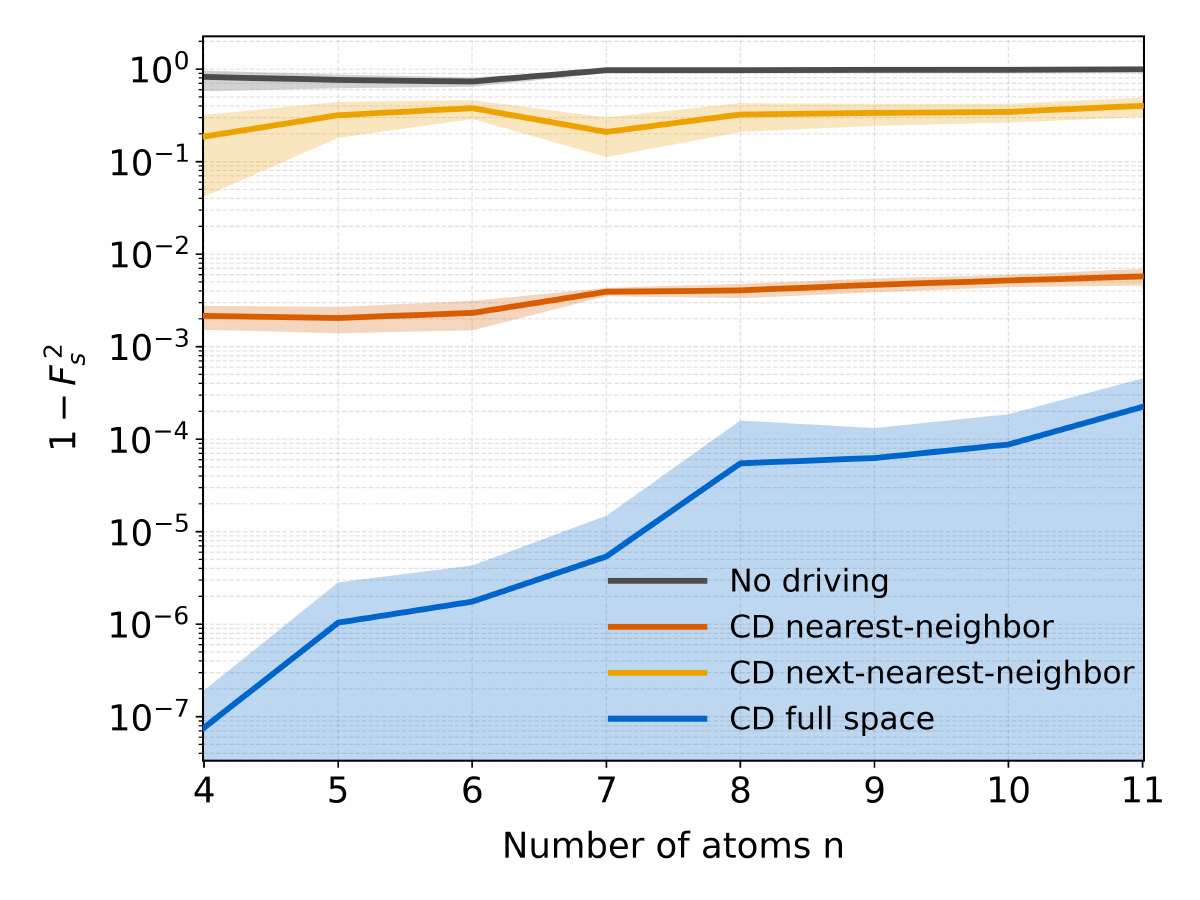}
        \vspace{-1.0em}
    \end{minipage}
    \hfill
    \begin{minipage}[t]{0.48\textwidth}
        \begin{picture}(0,0)
            \put(-136,-15){\textbf{(b)}}  % move label: left/right & up/down
        \end{picture}
        \includegraphics[width=\linewidth]{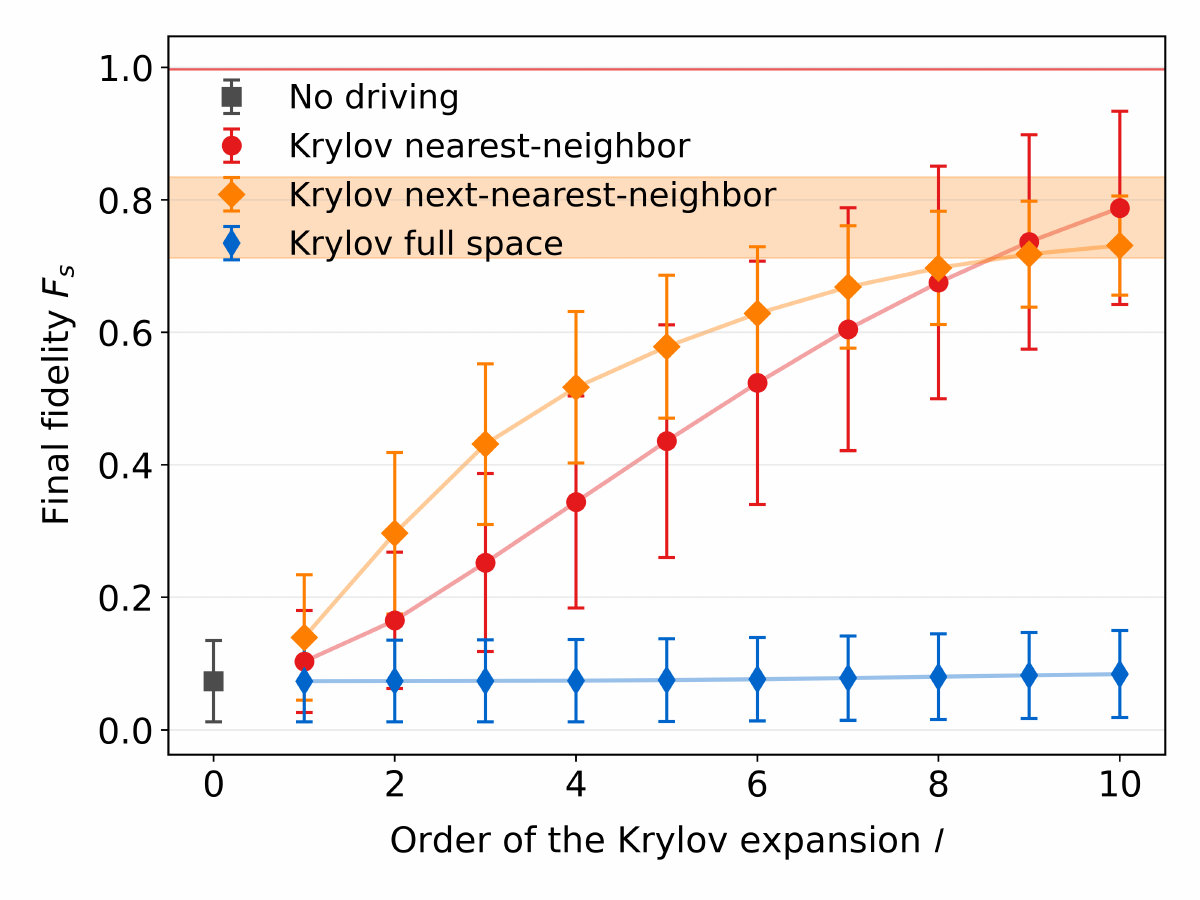}
        \vspace{-1.0em}
    \end{minipage}
    \vspace{-0.7em}
    \caption{\protect\justifying 
    (a) The error measure $1-F_{s}^2$ plotted on a logarithmic scale as a function of the number of atoms. The shaded bands indicate one standard deviation, and the solid line shows the average over all instances. The gray line gives the final fidelity without any counterdiabatic (CD) driving. The red and amber yellow lines show the fidelities obtained using the counterdiabatic terms derived from the subspace method with nearest-neighbor and next-nearest-neighbor subspaces, while the blue line shows the results from full space counterdiabatic driving. 
    (b) Final fidelity $F_{s}$ as a function of the order of the Krylov expansion (equivalently, the dimension of the Krylov space) $l$, evaluated over 150 distinct 11-atom configurations. Red and amber yellow markers indicate results from the Krylov method using submatrices restricted to nearest-neighbor and next-nearest-neighbor interactions. Results from the full space Krylov method and the no driving case are shown in blue and gray for comparison. Error bars denote one standard deviation. The red and amber yellow shaded regions indicate the upper bound achievable by the Krylov method, corresponding to the mean ± one standard deviation obtained from subspace diagonalization.
    }
    \label{fig:Su}
\end{figure*}

Final fidelity is used here as a performance metric to evaluate the effectiveness of the subspace approximation, as shown in Fig.~\ref{fig:Su} (a). We compare the resulting final fidelity obtained from time evolution with either only the time-dependent Hamiltonian $H_{0}(t)$ as stated in Eq.~\eqref{eq:H_ryd} (gray line) or Hamiltonian accompanied by counterdiabatic driving derived with the full space (blue line) or subspace protocol (red and amber yellow line).  While the full space counterdiabatic protocol represents the best possible performance achievable under the near-perfect numerical accuracy and perfect control, the red lines represent the final fidelity obtained from the time-dependent Hamiltonian with counterdiabatic driving term derived from solving the diagonalization in the $d_{IS}$-dimensional reduced subspaces where nearest-neighbor excitations are excluded. We can further simplify the protocol by excluding not only the nearest-neighbor but also the next-nearest-neighbor excitations, which makes the Hamiltonian in the subspace even smaller; the performance is shown in the amber yellow line. Compared to the no driving case, the nearest-neighbor subspace approximation counterdiabatic driving method provides an enhancement in the mean fidelity, ranging from 0.07-0.52 in the no-driving situation to 0.997-0.999. Even for the next-nearest-neighbor case, which is further simplified, fidelity remains above 0.77 for the number of atoms shown here.

For each number of atoms $n$, tens to hundreds of king's graphs (also called Union-Jack-type unit-disk graphs) are sampled and averaged. The filling fraction on the lattice graph is kept between 0.62 and 0.83 to ensure the problem remains non-trivial, avoiding both high-dropout cases and overly structured, lattice-like graphs that are easier to solve. Also, symmetry-equivalent configurations under rotation, inversion, translation, or combinations of these are counted only once in the statistical analysis. The time evolution is performed in full space to ensure accurate simulation of the systems.

\subsubsection{Krylov method}
Since the Krylov method is known to be computationally easier than exact diagonalization in finding the counterdiabatic matrix, we also evaluate the final fidelity obtained using a limited number of Krylov basis operators. We compare the final fidelity of the 11-atom cases using the Krylov basis operators derived from the full space (blue), nearest-neighbor subspace (red), and next-nearest-neighbor subspace (amber yellow) in Fig.~\ref{fig:Su}(b). While the reduced matrix method imposes a lower maximum achievable fidelity represented in the shaded region due to the smaller amount of information it uses, it converges toward this limit more quickly than the full-space method. It is particularly noteworthy that we are approximating a counterdiabatic operator with up to $O(4^{11})$ degrees of freedom, using only an $O(1)$ number of the Krylov basis operators. Remarkably, fidelities approaching 0.8 are achieved with just ten Krylov basis operators in the nearest-neighbor subspace, and above 0.7 in the next-nearest-neighbor case. Additionally, when only a small number of operators is allowed (e.g., 2–6 terms for the 11-atom case), the next-nearest-neighbor subspace often yields better fidelity, suggesting it may be a more efficient and computationally simpler choice in situations where only a limited number of operators is feasible, which is frequently the constraint in real experimental settings.
\begin{figure*}[t]
    \centering    
    \includegraphics[width=1.0\textwidth]{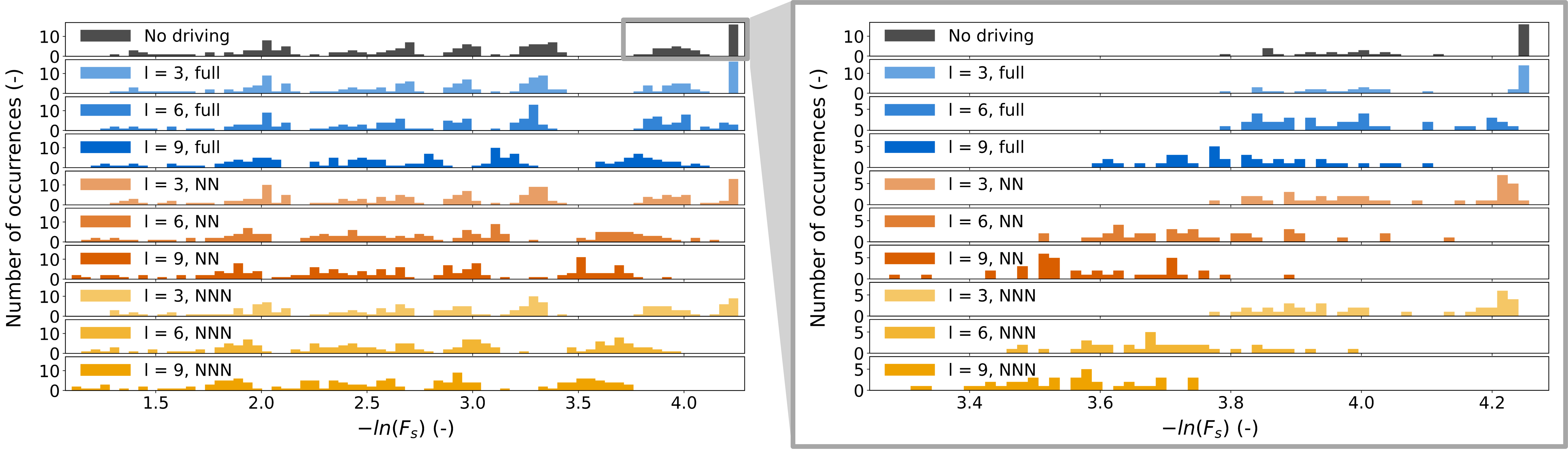}
    \vspace{-0.5em}
    \caption{\protect\justifying Distribution of occurrences as a function of the negative log final fidelity $-\ln(F_{s})$ for 11-atom configurations. The left panel shows statistics over 150 instances, while the right panel highlights the distribution for the hardest graph instances. Improvements are shown for three approaches: the standard full space Krylov method (blue), the Krylov method with a nearest-neighbor (NN) submatrix cost function (red), and the next-nearest-neighbor (NNN) submatrix variant (amber yellow). For each method, results are provided for the Krylov orders $l=3, 6, 9$.}
\label{fig:histogram}
\end{figure*}

\subsection{Subspace Cost Function}
In the previous section, the main purpose was to find the counterdiabatic matrix needed for a specific experiment in a computationally simpler way. In this section, instead of conducting the whole procedure from Eq.~\eqref{eq:A} to Eq.~\eqref{eq:Q} in the subspace, we attempt to compute only the the minimization of the cost function in Eq.~\eqref{eq:mincost} in the subspace. In the standard implementation of the Krylov method for approximating the exact counterdiabatic matrix, the cost function is typically defined as Eq.~\eqref{eq:Sl}. The optimization seeks to minimize the action $S_{l}$ by solving for the coefficients ${\{\alpha_{i}}\}$ in the expansion of the gauge potential in terms of the Krylov basis operators, as described in Eq.~\eqref{eq:Sl},~\eqref{eq:G},~\eqref{eq:dSl_da}.

Building on our earlier demonstration that subspace projections capture the dominant contributions of the counterdiabatic matrix, we propose a revised cost function based on subspace Frobenius products rather than those over the full Hilbert space. Specifically, we apply the optimization condition in Eq.~\eqref{eq:mincost}, but with the Frobenius inner product replaced by its subspace restricted form, $\left<A,B\right>_{F,sub}={\rm Tr}(A^{\dag}P_{IS}BP_{IS})$, where $P_{IS}$ is the projector onto the independent set subspace defined in Eq.~\eqref{eq:PIS}. This restriction allows us to focus the optimization on the relevant subspace. A small additional term proportional to $\epsilon\left<A,B\right>_{F}$ is included only to ensure numerical stability and to avoid minor overfitting or degeneracy issues, but it does not influence the optimization in any essential way. The major difference between this cost function and the commonly used one is that here we focus on the Frobenius product in the subspace instead of the full space, which places greater emphasis on the important part of the counterdiabatic matrix that we need to address.

Figure~\ref{fig:histogram} presents statistics for 11-atom systems over 150 random atom-graph instances, with final fidelity as the performance metric. We emphasize the most challenging MIS problems. While the original nested commutator approach (blue bars) yields some performance improvement, employing the subspace-based cost functions, including both the nearest-neighbor (red) and next-nearest-neighbor (amber yellow), consistently achieves superior results regardless of the order $l$ of the Krylov basis operators.

\subsection{Exploring Subspace Structure}
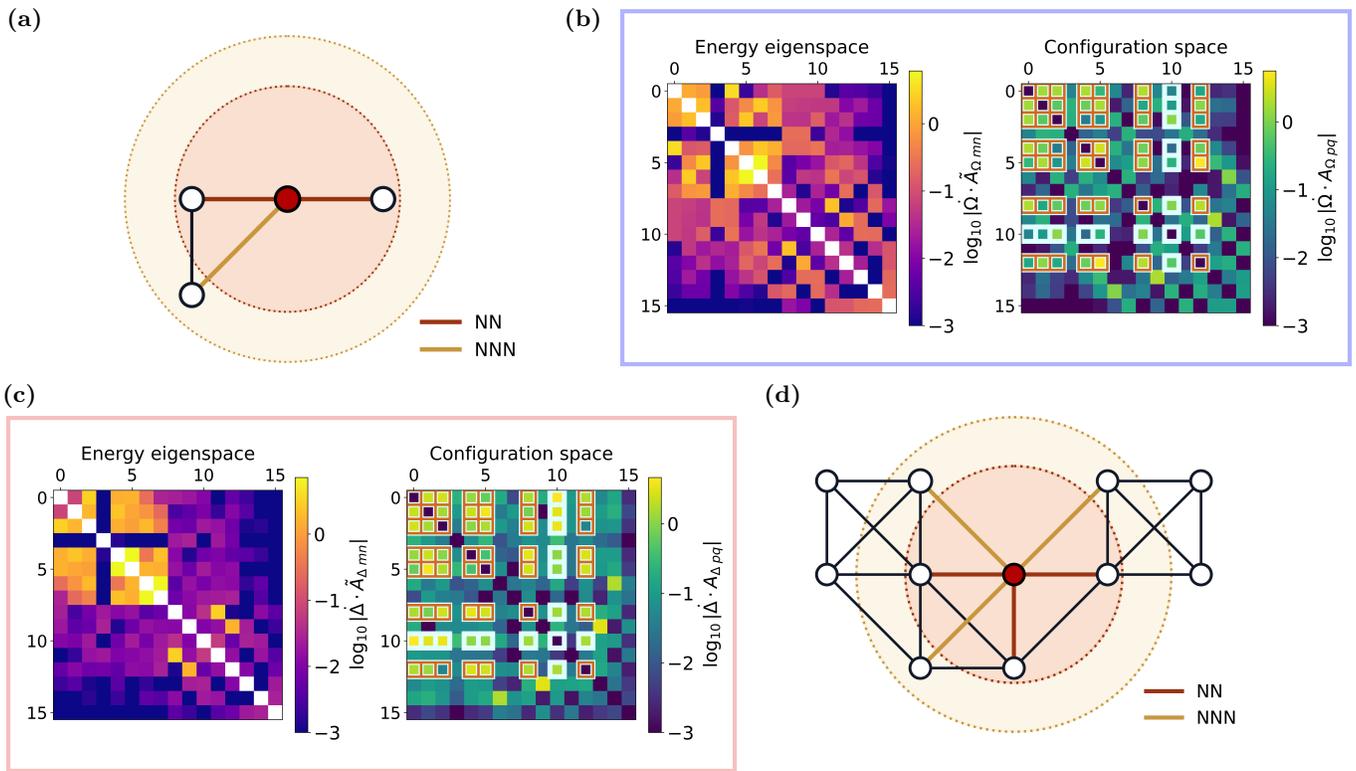
\begin{figure*}[htbp]
    \centering
    % Top row: A and smaller B
    \begin{minipage}[t]{0.5\textwidth}
        \raggedright
        \hspace{-1em}
        \begin{picture}(0,0)
            \put(-40,130){\textbf{(a)}}  % move label: left/right & up/down
        \end{picture}
        \includegraphics[height=0.50\linewidth]{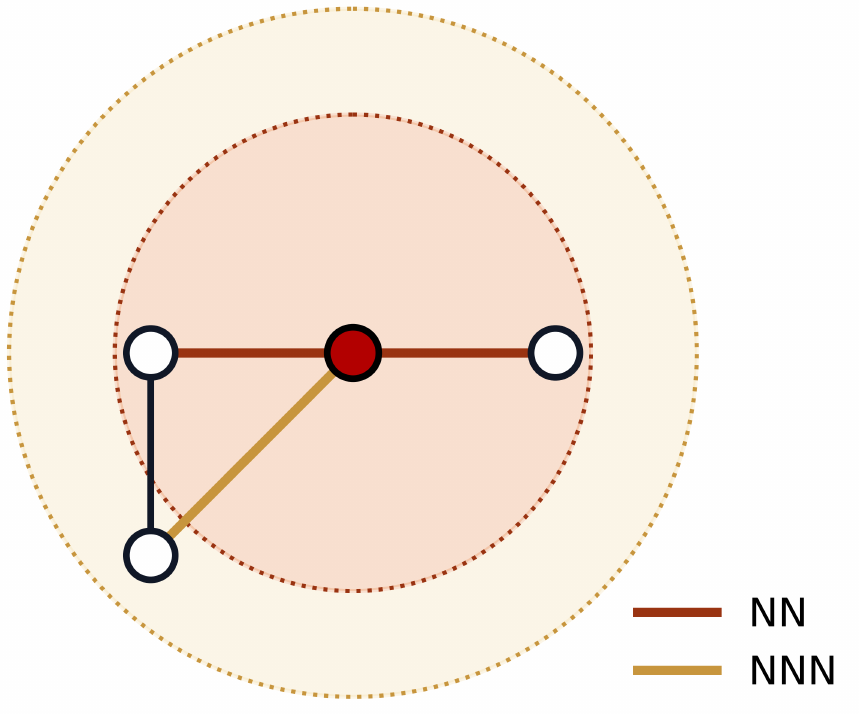}
    \end{minipage}
    \hspace{-0.15\textwidth}
    \begin{minipage}[t]{0.45\textwidth}
        \centering
        \begin{picture}(0,0)
            \put(-20,130){\textbf{(b)}}  % move label: left/right & up/down
        \end{picture}\includegraphics[height=0.60\linewidth]{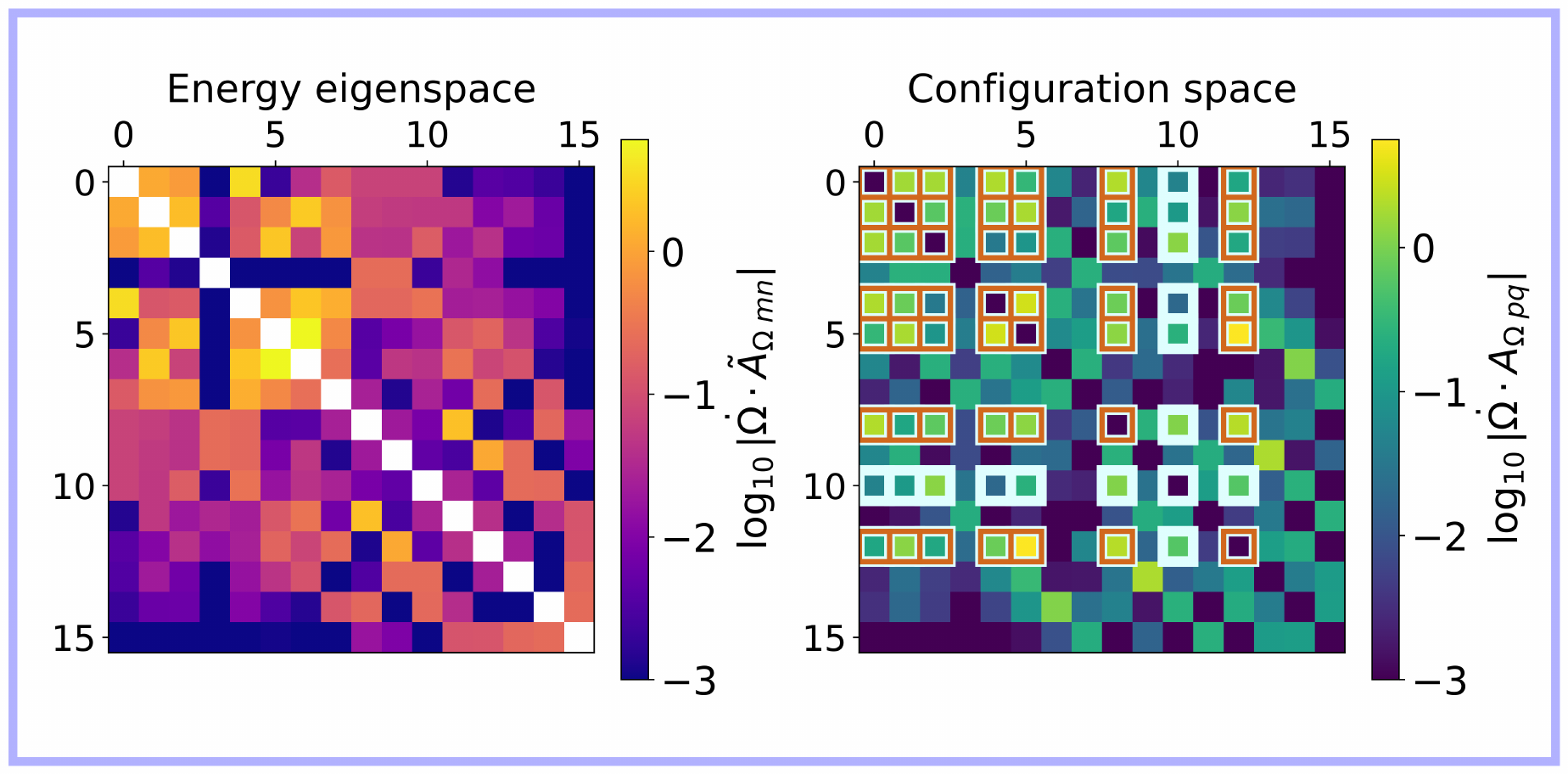}
    \end{minipage}
    \vspace{0.0em}
    % Bottom row: Enlarged C and D
    \begin{minipage}[t]{0.45\textwidth}
        \centering
        \begin{picture}(0,0)
            \put(-0,142){\textbf{(c)}}  % move label: left/right & up/down
            \end{picture}\includegraphics[height=0.60\linewidth]{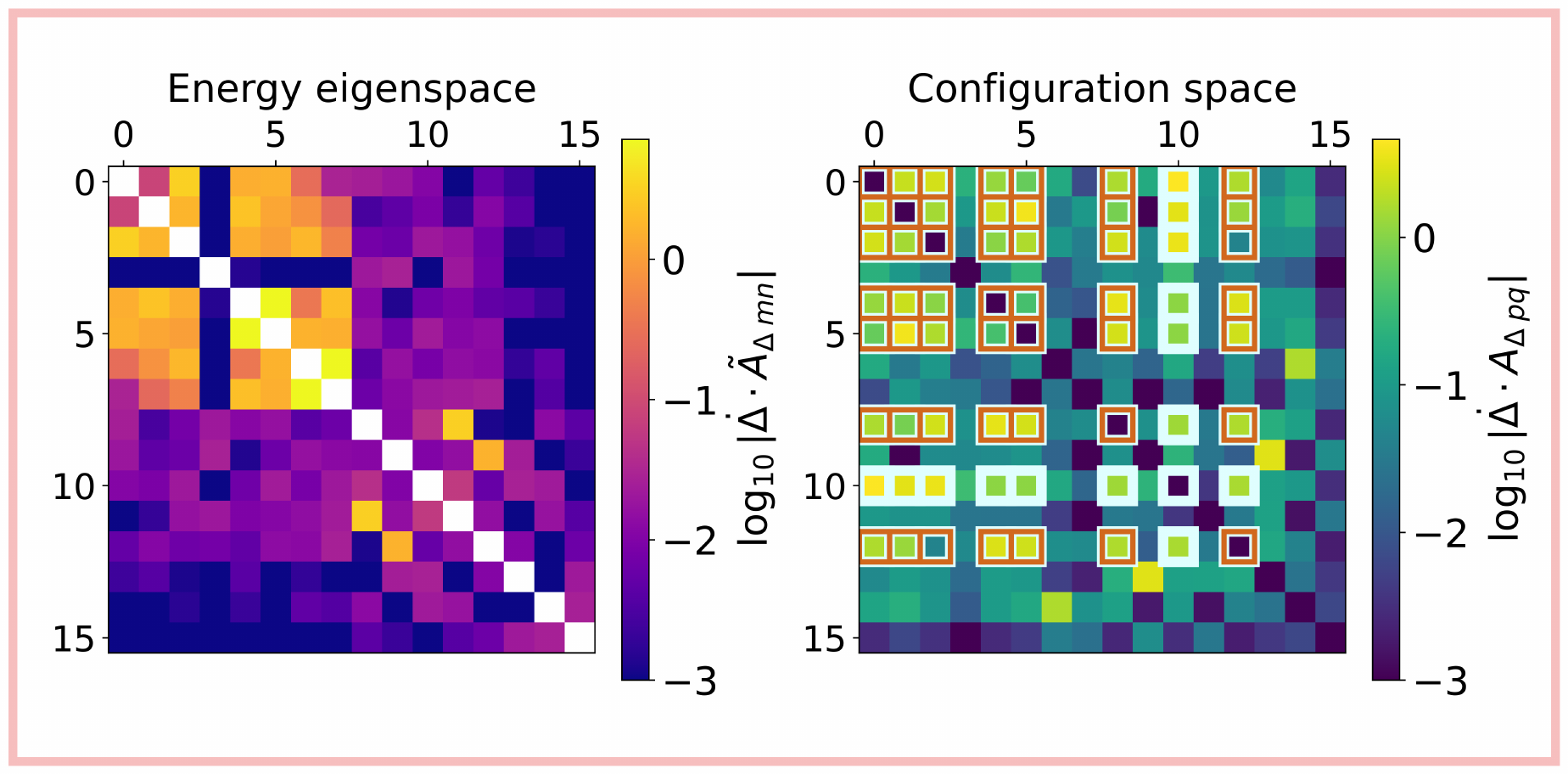}
    \end{minipage}
    \hfill
    \begin{minipage}[t]{0.5\textwidth}
        \centering
        \begin{picture}(15,-70)
            \put(-0,142){\textbf{(d)}}  % move label: left/right & up/down
        \end{picture}
        \includegraphics[height=0.60\linewidth]{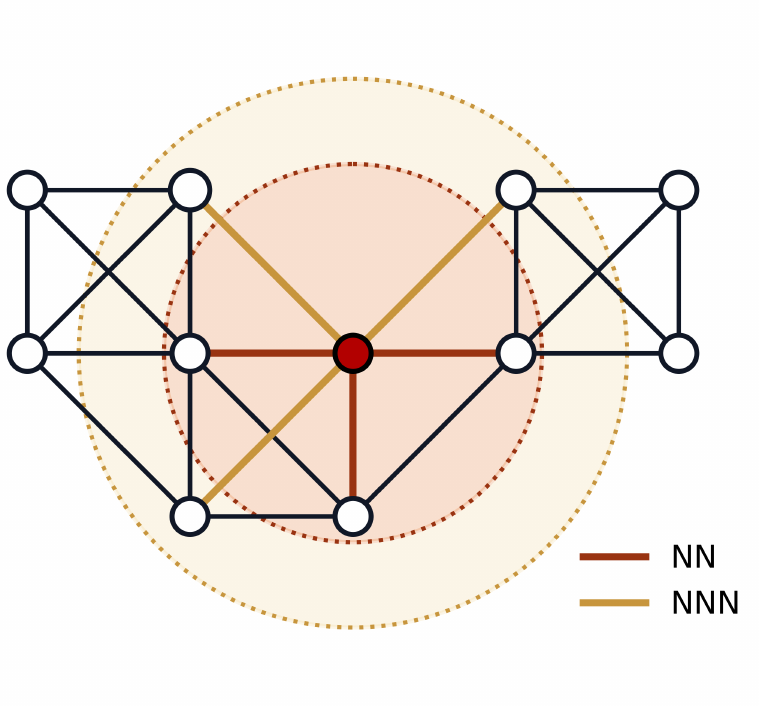}
    \end{minipage}
    \vspace{0.7em}
    \caption{\protect\justifying (a) Illustration of nearest-neighbor (NN) and next-nearest-neighbor (NNN) exclusions. For each atom, for example the selected red one in the graph, all the states involving the simultaneous excitation of two atoms connected by red edges are excluded from the counterdiabatic calculations. For the next-nearest-neighbor calculations, the simultaneous excitation of the atoms connected by red and amber yellow edges is excluded. This procedure is applied to each atom until all edges are considered. The large semi-transparent circles represent the radius to identify the nearest-neighbor (red) and next-nearest-neighbor (amber yellow) for the selected atom (red) in the graph. (b) Magnitude of elements of the Rabi-drive counterdiabatic term $\dot{\Omega}(t)A_{\Omega}(t)$ in energy and configuration bases, shown in log scale. In the energy eigenspace, each row and column represent an energy eigenstate ranked by their corresponding eigenvalues from low to high. In the configuration space, the elements selected by the nearest-neighbor subspace are marked by light cyan squares, and those selected by the next-nearest-neighbor subspace are marked by chocolate squares. (c) Same as panel (b), but for the detuning counterdiabatic term $\dot{\Delta}(t)A_\Delta(t)$. (d) An example of an 11-atom configuration with one atom highlighted to show its nearest-neighbor and next-nearest neighbor edges. As the number of atoms grows, the number of edges increases and further reduces the dimension of states in the subspace. 
    }
    \label{fig:CDmatrix}
\end{figure*}

For the time slices shown in Fig.~\ref{fig:IS}(b), the corresponding adiabatic gauge potential matrices for the four-atom configurations illustrated in Fig.~\labelcref{fig:CDmatrix}(a), scaled by the time derivatives of their respective parameters, $\dot{\Omega}(t)$ and $\dot{\Delta}(t)$, are plotted in Fig.~\labelcref{fig:CDmatrix}(b) and~\labelcref{fig:CDmatrix}(c). This scaling helps clarify the actual physical contributions of each component in the counterdiabatic protocol. For example, we show the Rabi-drive counterdiabatic term $\dot{\Omega}(t)A_{\Omega}(t)$ in Fig.~\labelcref{fig:CDmatrix}(b), where each matrix element is colored based on its magnitude after applying the absolute value and a base-10 logarithm. The left panels displays the Rabi-drive counterdiabatic term $\dot{\Omega}(t)A_{\Omega}(t)$ in the energy eigenspace. A key observation is that the largest matrix elements are concentrated along the diagonal, while the values become darker and weaker toward the off-diagonal edges. This suggests that transitions are primarily between states with similar energy. The right panel presents the same counterdiabatic matrix in the configuration basis. The light cyan and chocolate circled squares highlight matrix elements corresponding to configuration states without any simultaneously excited nearest-neighbor and next-nearest-neighbor pairs, respectively. These highlighted elements account for most of the largest values in the counterdiabatic matrix, which explains why the subspace method is particularly powerful.

This subspace approach offers two key advantages: not only does it lead to better fidelity, especially in hard instances, but it also significantly reduces computational complexity. Given that the total matrix size scales exponentially with the number of atoms, and computational tasks such as diagonalization scale cubically, this reduction represents a meaningful advantage for scaling to larger problem sizes. The application of the nearest-neighbor and next-nearest-neighbor subspaces highlights the potential for subspace-based counterdiabatic approaches in large-scale quantum optimization.

\section{Conclusion}
\label{Conclusion}
In conclusion, our study proposes a more efficient subspace method for identifying counterdiabatic driving strategies in solving the maximum independent set problem with Rydberg atom systems. One of the main challenges in applying counterdiabatic techniques is the complexity of constructing the driving matrix. Our results demonstrate that this difficulty can be significantly mitigated by diagonalizing a much smaller, carefully chosen submatrix that captures the essential elements of the full counterdiabatic matrix.

First, we show that this subspace approach can achieve high fidelity with substantially simpler diagonalization, increasing the probability of obtaining the MIS state from less than half to above $99.7\%$. Beyond exact diagonalization, we also demonstrate that this subspace approach can improve the Krylov method used for finding the gauge potential. In particular, combining the subspace method with the Krylov approach enables rapid convergence of the counterdiabatic drive. We show that both the diagonalization method and the Krylov method enhanced with graph-based subspaces provide efficient paths for constructing the counterdiabatic matrix. 

We also demonstrate that using a subspace-based cost function to target the most significant elements contributing to the counterdiabatic matrix can significantly improve the final fidelity compared to the conventional full-matrix Krylov method. In cases where experimental limitations constrain the number of implementable nested commutator terms (e.g., in Floquet engineering), larger reduced submatrices still offer improved performance.

Overall, our results underscore the practical promise of counterdiabatic driving in quantum annealing, while also highlighting some of the pitfalls in determining approximate driving schemes.

\emph{Acknowledgments}
The Flatiron Institute is a division of the Simons Foundation. D.S. and W.T.H. thank AFOSR for support through Award no. FA9550-25-1-0067 and NSF through Award no. 2105081.

\bibliographystyle{apsrev4-2}
\bibliography{hsieh2024}

@article{sels2017minimizing,
  title={Minimizing irreversible losses in quantum systems by local counterdiabatic driving},
  author={Sels, Dries and Polkovnikov, Anatoli},
  journal={Proceedings of the National Academy of Sciences},
  volume={114},
  number={20},
  pages={E3909--E3916},
  year={2017},
  publisher={National Acad Sciences}
}

@article{claeys2019floquet,
  title={Floquet-engineering counterdiabatic protocols in quantum many-body systems},
  author={Claeys, Pieter W and Pandey, Mohit and Sels, Dries and Polkovnikov, Anatoli},
  journal={Physical review letters},
  volume={123},
  number={9},
  pages={090602},
  year={2019},
  publisher={APS}
}

@book{mcgeoch2014adiabatic,
  title={Adiabatic quantum computation and quantum annealing: Theory and practice},
  author={McGeoch, Catherine C},
  year={2014},
  publisher={Morgan \& Claypool Publishers}
}

@article{albash2018adiabatic,
  title={Adiabatic quantum computation},
  author={Albash, Tameem and Lidar, Daniel A},
  journal={Reviews of Modern Physics},
  volume={90},
  number={1},
  pages={015002},
  year={2018},
  publisher={APS}
}

@article{kolodrubetz2017geometry,
  title={Geometry and non-adiabatic response in quantum and classical systems},
  author={Kolodrubetz, Michael and Sels, Dries and Mehta, Pankaj and Polkovnikov, Anatoli},
  journal={Physics Reports},
  volume={697},
  pages={1--87},
  year={2017},
  publisher={Elsevier}
}

@article{vcepaite2023counterdiabatic,
  title={Counterdiabatic optimized local driving},
  author={{\v{C}}epait{\.e}, Ieva and Polkovnikov, Anatoli and Daley, Andrew J and Duncan, Callum W},
  journal={PRX Quantum},
  volume={4},
  number={1},
  pages={010312},
  year={2023},
  publisher={APS}
}

@article{morawetz2024efficient,
  title={Efficient paths for local counterdiabatic driving},
  author={Morawetz, Stewart and Polkovnikov, Anatoli},
  journal={Physical Review B},
  volume={110},
  number={2},
  pages={024304},
  year={2024},
  publisher={APS}
}

@article{del2013shortcuts,
  title={Shortcuts to adiabaticity by counterdiabatic driving},
  author={Del Campo, Adolfo},
  journal={Physical review letters},
  volume={111},
  number={10},
  pages={100502},
  year={2013},
  publisher={APS}
}

@article{duncan2024exact,
  title={Exact counterdiabatic driving in finite topological lattice models},
  author={Duncan, Callum W},
  journal={Physical Review B},
  volume={109},
  number={24},
  pages={245421},
  year={2024},
  publisher={APS}
}

@article{bukov2019geometric,
  title={Geometric speed limit of accessible many-body state preparation},
  author={Bukov, Marin and Sels, Dries and Polkovnikov, Anatoli},
  journal={Physical Review X},
  volume={9},
  number={1},
  pages={011034},
  year={2019},
  publisher={APS}
}

@article{nakahara2022counterdiabatic,
  title={Counterdiabatic formalism of shortcuts to adiabaticity},
  author={Nakahara, Mikio},
  journal={Philosophical Transactions of the Royal Society A},
  volume={380},
  number={2239},
  pages={20210272},
  year={2022},
  publisher={The Royal Society}
}

@article{yague2023shortcut,
  title={Shortcut-to-Adiabatic Controlled-Phase Gate in Rydberg Atoms},
  author={Yag{\"u}e Bosch, Luis S and Ehret, Tim and Petiziol, Francesco and Arimondo, Ennio and Wimberger, Sandro},
  journal={Annalen der Physik},
  volume={535},
  number={12},
  pages={2300275},
  year={2023},
  publisher={Wiley Online Library}
}

@article{ebadi2022quantum,
  title={Quantum optimization of maximum independent set using Rydberg atom arrays},
  author={Ebadi, Sepehr and Keesling, Alexander and Cain, Madelyn and Wang, Tout T and Levine, Harry and Bluvstein, Dolev and Semeghini, Giulia and Omran, Ahmed and Liu, J-G and Samajdar, Rhine and others},
  journal={Science},
  volume={376},
  number={6598},
  pages={1209--1215},
  year={2022},
  publisher={American Association for the Advancement of Science}
}

@article{saberi14,
  title = {Adiabatic tracking of quantum many-body dynamics},
  author = {Saberi, Hamed and Opatrn\'y, Tom\'a\ifmmode \check{s}\else \v{s}\fi{} and M\o{}lmer, Klaus and del Campo, Adolfo},
  journal = {Phys. Rev. A},
  volume = {90},
  issue = {6},
  pages = {060301},
  numpages = {5},
  year = {2014},
  month = {Dec},
  publisher = {American Physical Society},
  doi = {10.1103/PhysRevA.90.060301},
  url = {https://link.aps.org/doi/10.1103/PhysRevA.90.060301}
}

@article{finzgar25,
  title = {Counterdiabatic Driving with Performance Guarantees},
  author = {Fin\ifmmode \check{z}\else \v{z}\fi{}gar, Jernej Rudi and Notarnicola, Simone and Cain, Madelyn and Lukin, Mikhail D. and Sels, Dries},
  journal = {Phys. Rev. Lett.},
  volume = {135},
  issue = {18},
  pages = {180602},
  numpages = {9},
  year = {2025},
  month = {Oct},
  publisher = {American Physical Society},
  doi = {10.1103/pqhl-nbtk},
  url = {https://link.aps.org/doi/10.1103/pqhl-nbtk}
}

@article{morawetz2025universal,
  title={Universal counterdiabatic driving in Krylov space},
  author={Morawetz, Stewart and Polkovnikov, Anatoli},
  journal={PRX Quantum},
  volume={6},
  number={4},
  pages={040320},
  year={2025},
  publisher={APS}
}

@article{hegade2022digitized,
  title={Digitized counterdiabatic quantum optimization},
  author={Hegade, Narendra N and Chen, Xi and Solano, Enrique},
  journal={Physical Review Research},
  volume={4},
  number={4},
  pages={L042030},
  year={2022},
  publisher={APS}
}

@article{takahashi2024shortcuts,
  title={Shortcuts to adiabaticity in Krylov space},
  author={Takahashi, Kazutaka and del Campo, Adolfo},
  journal={Physical Review X},
  volume={14},
  number={1},
  pages={011032},
  year={2024},
  publisher={APS}
}

@article{zhang2024analog,
  title={Analog Counterdiabatic Quantum Computing},
  author={Zhang, Qi and Hegade, Narendra N and Cadavid, Alejandro Gomez and Lassabli{\`e}re, Lucas and Trautmann, Jan and Perseguers, S{\'e}bastien and Solano, Enrique and Henriet, Lo{\"\i}c and Michon, Eric},
  journal={arXiv preprint arXiv:2405.14829},
  year={2024}
}

@article{lukin2024quantum,
  title={Quantum quench dynamics as a shortcut to adiabaticity},
  author={Lukin, Alexander and Schiffer, Benjamin F and Braverman, Boris and Cantu, Sergio H and Huber, Florian and Bylinskii, Alexei and Amato-Grill, Jesse and Maskara, Nishad and Cain, Madelyn and Wild, Dominik S and others},
  journal={arXiv preprint arXiv:2405.21019},
  year={2024}
}

@article{tseng2013counterdiabatic,
  title={Counterdiabatic mode-evolution based coupled-waveguide devices},
  author={Tseng, Shuo-Yen},
  journal={Optics Express},
  volume={21},
  number={18},
  pages={21224--21235},
  year={2013},
  publisher={Optica Publishing Group}
}

@article{prielinger2021two,
  title={Two-parameter counter-diabatic driving in quantum annealing},
  author={Prielinger, Luise and Hartmann, Andreas and Yamashiro, Yu and Nishimura, Kohji and Lechner, Wolfgang and Nishimori, Hidetoshi},
  journal={Physical Review Research},
  volume={3},
  number={1},
  pages={013227},
  year={2021},
  publisher={APS}
}

@article{chung2019shortcuts,
  title={Shortcuts to adiabaticity in optical waveguides},
  author={Chung, H-C and Mart{\'\i}nez-Garaot, S and Chen, X and Muga, JG and Tseng, S-Y},
  journal={Europhysics Letters},
  volume={127},
  number={3},
  pages={34001},
  year={2019},
  publisher={IOP Publishing}
}

@article{gangopadhay2025counterdiabatic,
  title={Counterdiabatic route to entanglement steering and dynamical freezing in the Floquet Lipkin-Meshkov-Glick model},
  author={Gangopadhay, Nakshatra and Choudhury, Sayan},
  journal={Physical Review Letters},
  volume={135},
  number={2},
  pages={020407},
  year={2025},
  publisher={APS}
}

@article{farhi2000quantum,
  title={Quantum computation by adiabatic evolution},
  author={Farhi, Edward and Goldstone, Jeffrey and Gutmann, Sam and Sipser, Michael},
  journal={arXiv preprint quant-ph/0001106},
  year={2000}
}

@article{kim2024variational,
  title={Variational adiabatic transport of tensor networks},
  author={Kim, Hyeongjin and Fishman, Matthew and Sels, Dries},
  journal={PRX Quantum},
  volume={5},
  number={2},
  pages={020361},
  year={2024},
  publisher={APS}
}

@article{shende2024experimental,
  title={Experimental investigation of a quantum Otto heat engine with shortcuts to adiabaticity implemented using counter-adiabatic driving},
  author={Shende, Krishna and Kandpal, Matreyee and Dorai, Kavita and others},
  journal={arXiv preprint arXiv:2412.20194},
  year={2024}
}

@article{pichler2018quantum,
  title={Quantum optimization for maximum independent set using Rydberg atom arrays},
  author={Pichler, Hannes and Wang, Sheng-Tao and Zhou, Leo and Choi, Soonwon and Lukin, Mikhail D},
  journal={arXiv preprint arXiv:1808.10816},
  year={2018}
}

@article{bottarelli2024symmetry,
  title={Symmetry-enhanced Counterdiabatic Quantum Algorithm for Qudits},
  author={Bottarelli, Alberto and de Andoin, Mikel Garcia and Chandarana, Pranav and Paul, Koushik and Chen, Xi and Sanz, Mikel and Hauke, Philipp},
  journal={arXiv preprint arXiv:2410.06710},
  year={2024}
}

@article{vizzuso2024role,
  title={The role of gaps in digitized counterdiabatic QAOA for fully-connected spin models},
  author={Vizzuso, Mara and Passarelli, Gianluca and Cantele, Giovanni and Lucignano, Procolo},
  journal={arXiv preprint arXiv:2409.03503},
  year={2024}
}

@article{andrist2023hardness,
  title={Hardness of the maximum-independent-set problem on unit-disk graphs and prospects for quantum speedups},
  author={Andrist, Ruben S and Schuetz, Martin JA and Minssen, Pierre and Yalovetzky, Romina and Chakrabarti, Shouvanik and Herman, Dylan and Kumar, Niraj and Salton, Grant and Shaydulin, Ruslan and Sun, Yue and others},
  journal={Physical Review Research},
  volume={5},
  number={4},
  pages={043277},
  year={2023},
  publisher={APS}
}

@misc{bloqade2023quera,
  url = {https://github.com/QuEraComputing/Bloqade.jl/},
  title = {Bloqade.jl: {P}ackage for the quantum computation and quantum simulation based on the neutral-atom architecture.},
  year = {2023}
}

@article{li2024quantum,
  title={Quantum counterdiabatic driving with local control},
  author={Li, Changhao and Shen, Jiayu and Shaydulin, Ruslan and Pistoia, Marco},
  journal={arXiv preprint arXiv:2403.01854},
  year={2024}
}

@article{visuri2025digitized,
  title={Digitized counterdiabatic quantum critical dynamics},
  author={Visuri, Anne-Maria and Cadavid, Alejandro Gomez and Bhargava, Balaganchi A and Romero, Sebasti{\'a}n V and Grabarits, Andr{\'a}s and Chandarana, Pranav and Solano, Enrique and del Campo, Adolfo and Hegade, Narendra N},
  journal={arXiv preprint arXiv:2502.15100},
  year={2025}
}

@article{hatomura2024shortcuts,
  title={Shortcuts to adiabaticity: theoretical framework, relations between different methods, and versatile approximations},
  author={Hatomura, Takuya},
  journal={Journal of Physics B: Atomic, Molecular and Optical Physics},
  volume={57},
  number={10},
  pages={102001},
  year={2024},
  publisher={IOP Publishing}
}

@article{berry2009transitionless,
  title={Transitionless quantum driving},
  author={Berry, Michael Victor},
  journal={Journal of Physics A: Mathematical and Theoretical},
  volume={42},
  number={36},
  pages={365303},
  year={2009},
  publisher={IOP Publishing}
}

@article{pandey2020adiabatic,
  title={Adiabatic eigenstate deformations as a sensitive probe for quantum chaos},
  author={Pandey, Mohit and Claeys, Pieter W and Campbell, David K and Polkovnikov, Anatoli and Sels, Dries},
  journal={Physical Review X},
  volume={10},
  number={4},
  pages={041017},
  year={2020},
  publisher={APS}
}

@article{passarelli2020counterdiabatic,
  title={Counterdiabatic driving in the quantum annealing of the p-spin model: A variational approach},
  author={Passarelli, Gianluca and Cataudella, Vittorio and Fazio, Rosario and Lucignano, Procolo},
  journal={Physical Review Research},
  volume={2},
  number={1},
  pages={013283},
  year={2020},
  publisher={APS}
}

@article{aharonov2008adiabatic,
  title={Adiabatic quantum computation is equivalent to standard quantum computation},
  author={Aharonov, Dorit and Van Dam, Wim and Kempe, Julia and Landau, Zeph and Lloyd, Seth and Regev, Oded},
  journal={SIAM review},
  volume={50},
  number={4},
  pages={755--787},
  year={2008},
  publisher={SIAM}
}

@incollection{karp2009reducibility,
  title={Reducibility among combinatorial problems},
  author={Karp, Richard M},
  booktitle={50 Years of Integer Programming 1958-2008: from the Early Years to the State-of-the-Art},
  pages={219--241},
  year={2009},
  publisher={Springer}
}

@article{childs2000finding,
  title={Finding cliques by quantum adiabatic evolution},
  author={Childs, Andrew M and Farhi, Edward and Goldstone, Jeffrey and Gutmann, Sam},
  journal={arXiv preprint quant-ph/0012104},
  year={2000}
}

@article{farhi2001quantum,
  title={A quantum adiabatic evolution algorithm applied to random instances of an NP-complete problem},
  author={Farhi, Edward and Goldstone, Jeffrey and Gutmann, Sam and Lapan, Joshua and Lundgren, Andrew and Preda, Daniel},
  journal={Science},
  volume={292},
  number={5516},
  pages={472--475},
  year={2001},
  publisher={American Association for the Advancement of Science}
}

@article{kadowaki1998quantum,
  title={Quantum annealing in the transverse Ising model},
  author={Kadowaki, Tadashi and Nishimori, Hidetoshi},
  journal={Physical Review E},
  volume={58},
  number={5},
  pages={5355},
  year={1998},
  publisher={APS}
}

@article{weinberg2020scaling,
  title={Scaling and diabatic effects in quantum annealing with a D-Wave device},
  author={Weinberg, Phillip and Tylutki, Marek and R{\"o}nkk{\"o}, Jami M and Westerholm, Jan and {\AA}str{\"o}m, Jan A and Manninen, Pekka and T{\"o}rm{\"a}, P{\"a}ivi and Sandvik, Anders W},
  journal={Physical Review Letters},
  volume={124},
  number={9},
  pages={090502},
  year={2020},
  publisher={APS}
}

@article{vinci2016nested,
  title={Nested quantum annealing correction},
  author={Vinci, Walter and Albash, Tameem and Lidar, Daniel A},
  journal={npj Quantum Information},
  volume={2},
  number={1},
  pages={1--6},
  year={2016},
  publisher={Nature Publishing Group}
}

@article{johnson2011quantum,
  title={Quantum annealing with manufactured spins},
  author={Johnson, Mark W and Amin, Mohammad HS and Gildert, Suzanne and Lanting, Trevor and Hamze, Firas and Dickson, Neil and Harris, Richard and Berkley, Andrew J and Johansson, Jan and Bunyk, Paul and others},
  journal={Nature},
  volume={473},
  number={7346},
  pages={194--198},
  year={2011},
  publisher={Nature Publishing Group UK London}
}

@article{lucas2014ising,
  title={Ising formulations of many NP problems},
  author={Lucas, Andrew},
  journal={Frontiers in physics},
  volume={2},
  pages={5},
  year={2014},
  publisher={Frontiers Media SA}
}

@article{king2022coherent,
  title={Coherent quantum annealing in a programmable 2,000 qubit Ising chain},
  author={King, Andrew D and Suzuki, Sei and Raymond, Jack and Zucca, Alex and Lanting, Trevor and Altomare, Fabio and Berkley, Andrew J and Ejtemaee, Sara and Hoskinson, Emile and Huang, Shuiyuan and others},
  journal={Nature Physics},
  volume={18},
  number={11},
  pages={1324--1328},
  year={2022},
  publisher={Nature Publishing Group UK London}
}

@article{hogg2003adiabatic,
  title={Adiabatic quantum computing for random satisfiability problems},
  author={Hogg, Tad},
  journal={Physical Review A},
  volume={67},
  number={2},
  pages={022314},
  year={2003},
  publisher={APS}
}

@article{altshuler2010anderson,
  title={Anderson localization makes adiabatic quantum optimization fail},
  author={Altshuler, Boris and Krovi, Hari and Roland, J{\'e}r{\'e}mie},
  journal={Proceedings of the National Academy of Sciences},
  volume={107},
  number={28},
  pages={12446--12450},
  year={2010},
  publisher={National Academy of Sciences}
}

@article{hegade2025digitized,
  title={Digitized Counterdiabatic Quantum Sampling},
  author={Hegade, Narendra N and Kortikar, Nachiket L and Bhargava, Balaganchi A and Hern{\'a}ndez, Juan FR and Cadavid, Alejandro Gomez and Chandarana, Pranav and Romero, Sebasti{\'a}n V and Kumar, Shubham and Simen, Anton and Visuri, Anne-Maria and others},
  journal={arXiv preprint arXiv:2510.26735},
  year={2025}
}

@article{schuetz2025qredumis,
  title={qReduMIS: A Quantum-Informed Reduction Algorithm for the Maximum Independent Set Problem},
  author={Schuetz, Martin JA and Yalovetzky, Romina and Andrist, Ruben S and Salton, Grant and Sun, Yue and Raymond, Rudy and Chakrabarti, Shouvanik and Acharya, Atithi and Shaydulin, Ruslan and Pistoia, Marco and others},
  journal={arXiv preprint arXiv:2503.12551},
  year={2025}
}

@article{bombieri2025quantum,
  title={Quantum adiabatic optimization with Rydberg arrays: localization phenomena and encoding strategies},
  author={Bombieri, Lisa and Zeng, Zhongda and Tricarico, Roberto and Lin, Rui and Notarnicola, Simone and Cain, Madelyn and Lukin, Mikhail D and Pichler, Hannes},
  journal={PRX Quantum},
  volume={6},
  number={2},
  pages={020306},
  year={2025},
  publisher={APS}
}

@inproceedings{sohrabizadeh2024gnn,
  title={GNN-Based Performance Prediction of Quantum Optimization of Maximum Independent Set},
  author={Sohrabizadeh, Atefeh and Lin, Wan-Hsuan and Tan, Daniel Bochen and Cain, Madelyn and Wang, Sheng-Tao and Lukin, Mikhail D and Cong, Jason},
  booktitle={Proceedings of the 43rd IEEE/ACM International Conference on Computer-Aided Design},
  pages={1--6},
  year={2024}
}

@article{nguyen2023quantum,
  title={Quantum optimization with arbitrary connectivity using Rydberg atom arrays},
  author={Nguyen, Minh-Thi and Liu, Jin-Guo and Wurtz, Jonathan and Lukin, Mikhail D and Wang, Sheng-Tao and Pichler, Hannes},
  journal={PRX Quantum},
  volume={4},
  number={1},
  pages={010316},
  year={2023},
  publisher={APS}
}

@article{ebadi2021quantum,
  title={Quantum phases of matter on a 256-atom programmable quantum simulator},
  author={Ebadi, Sepehr and Wang, Tout T and Levine, Harry and Keesling, Alexander and Semeghini, Giulia and Omran, Ahmed and Bluvstein, Dolev and Samajdar, Rhine and Pichler, Hannes and Ho, Wen Wei and others},
  journal={Nature},
  volume={595},
  number={7866},
  pages={227--232},
  year={2021},
  publisher={Nature Publishing Group UK London}
}

@article{manovitz2025quantum,
  title={Quantum coarsening and collective dynamics on a programmable simulator},
  author={Manovitz, Tom and Li, Sophie H and Ebadi, Sepehr and Samajdar, Rhine and Geim, Alexandra A and Evered, Simon J and Bluvstein, Dolev and Zhou, Hengyun and Koyluoglu, Nazli Ugur and Feldmeier, Johannes and others},
  journal={Nature},
  volume={638},
  number={8049},
  pages={86--92},
  year={2025},
  publisher={Nature Publishing Group UK London}
}

@article{trummer2015multiple,
  title={Multiple query optimization on the D-Wave 2X adiabatic quantum computer},
  author={Trummer, Immanuel and Koch, Christoph},
  journal={arXiv preprint arXiv:1510.06437},
  year={2015}
}

@article{king2025beyond,
  title={Beyond-classical computation in quantum simulation},
  author={King, Andrew D and Nocera, Alberto and Rams, Marek M and Dziarmaga, Jacek and Wiersema, Roeland and Bernoudy, William and Raymond, Jack and Kaushal, Nitin and Heinsdorf, Niclas and Harris, Richard and others},
  journal={Science},
  volume={388},
  number={6743},
  pages={199--204},
  year={2025},
  publisher={American Association for the Advancement of Science}
}

@article{orquin2025analog,
  title={Analog Quantum Feature Selection with Neutral-Atom Quantum Processors},
  author={Orquin-Marques, Jose J and Flores-Garrigos, Carlos and Cadavid, Alejandro Gomez and Simen, Anton and Solano, Enrique and Hegade, Narendra N and Martin-Guerrero, Jose D and Vives-Gilabert, Yolanda},
  journal={arXiv preprint arXiv:2510.20798},
  year={2025}
}

@article{simen2025digitized,
  title={Digitized Counterdiabatic Quantum Feature Extraction},
  author={Simen, Anton and Flores-Garrig{\'o}s, Carlos and De Oliveira, Murilo Henrique and Barrios, Gabriel Dario Alvarado and Cadavid, Alejandro Gomez and Dalal, Archismita and Solano, Enrique and Hegade, Narendra N and Zhang, Qi},
  journal={arXiv preprint arXiv:2510.13807},
  year={2025}
}

@article{romero2025bias,
  title={Bias-field digitized counterdiabatic quantum algorithm for higher-order binary optimization},
  author={Romero, Sebasti{\'a}n V and Visuri, Anne-Maria and Cadavid, Alejandro Gomez and Simen, Anton and Solano, Enrique and Hegade, Narendra N},
  journal={Communications Physics},
  volume={8},
  number={1},
  pages={348},
  year={2025},
  publisher={Nature Publishing Group UK London}
}

@article{chandarana2022digitized,
  title={Digitized-counterdiabatic quantum approximate optimization algorithm},
  author={Chandarana, Pranav and Hegade, Narendra N and Paul, Koushik and Albarr{\'a}n-Arriagada, Francisco and Solano, Enrique and Del Campo, Adolfo and Chen, Xi},
  journal={Physical review research},
  volume={4},
  number={1},
  pages={013141},
  year={2022},
  publisher={APS}
}

@article{hegade2021shortcuts,
  title={Shortcuts to adiabaticity in digitized adiabatic quantum computing},
  author={Hegade, Narendra N and Paul, Koushik and Ding, Yongcheng and Sanz, Mikel and Albarr{\'a}n-Arriagada, Francisco and Solano, Enrique and Chen, Xi},
  journal={Physical Review Applied},
  volume={15},
  number={2},
  pages={024038},
  year={2021},
  publisher={APS}
}

@article{demirplak2003adiabatic,
  title={Adiabatic population transfer with control fields},
  author={Demirplak, Mustafa and Rice, Stuart A},
  journal={The Journal of Physical Chemistry A},
  volume={107},
  number={46},
  pages={9937--9945},
  year={2003},
  publisher={ACS Publications}
}

@article{demirplak2005assisted,
  title={Assisted adiabatic passage revisited},
  author={Demirplak, Mustafa and Rice, Stuart A},
  journal={The Journal of Physical Chemistry B},
  volume={109},
  number={14},
  pages={6838--6844},
  year={2005},
  publisher={ACS Publications}
}

@article{van2024gate,
  title={Gate-based counterdiabatic driving with complexity guarantees},
  author={van Vreumingen, Dyon},
  journal={Physical Review A},
  volume={110},
  number={5},
  pages={052419},
  year={2024},
  publisher={APS}
}

@article{guery2019shortcuts,
  title={Shortcuts to adiabaticity: Concepts, methods, and applications},
  author={Gu{\'e}ry-Odelin, David and Ruschhaupt, Andreas and Kiely, Anthony and Torrontegui, Erik and Mart{\'\i}nez-Garaot, Sofia and Muga, Juan Gonzalo},
  journal={Reviews of Modern Physics},
  volume={91},
  number={4},
  pages={045001},
  year={2019},
  publisher={APS}
}

@article{del2012assisted,
  title={Assisted Finite-Rate Adiabatic Passage Across a Quantum Critical Point: Exact Solution for the Quantum Ising Model},
  author={del Campo, Adolfo and Rams, Marek M and Zurek, Wojciech H},
  journal={Physical review letters},
  volume={109},
  number={11},
  pages={115703},
  year={2012},
  publisher={APS}
}

@incollection{TORRONTEGUI2013117,
title = {Chapter 2 - Shortcuts to Adiabaticity},
editor = {Ennio Arimondo and Paul R. Berman and Chun C. Lin},
series = {Advances In Atomic, Molecular, and Optical Physics},
publisher = {Academic Press},
volume = {62},
pages = {117-169},
year = {2013},
booktitle = {Advances in Atomic, Molecular, and Optical Physics},
issn = {1049-250X},
doi = {https://doi.org/10.1016/B978-0-12-408090-4.00002-5},
url = {https://www.sciencedirect.com/science/article/pii/B9780124080904000025},
author = {Erik Torrontegui and Sara Ibáñez and Sofia Martínez-Garaot and Michele Modugno and Adolfo {del Campo} and David Guéry-Odelin and Andreas Ruschhaupt and Xi Chen and Juan Gonzalo Muga}
}

@article{schindler2024counterdiabatic,
  title={Counterdiabatic driving for periodically driven systems},
  author={Schindler, Paul M and Bukov, Marin},
  journal={Physical Review Letters},
  volume={133},
  number={12},
  pages={123402},
  year={2024},
  publisher={APS}
}

@article{beterov2025counterdiabatic,
  title={Counterdiabatic driving at Rydberg excitation for symmetric $ C\_Z $ gates with ultracold neutral atoms},
  author={Beterov, II and Kozenko, KV and Xu, P and Ryabtsev, II},
  journal={arXiv preprint arXiv:2510.04766},
  year={2025}
}

@article{zheng2016cost,
  title={Cost of counterdiabatic driving and work output},
  author={Zheng, Yuanjian and Campbell, Steve and De Chiara, Gabriele and Poletti, Dario},
  journal={Physical Review A},
  volume={94},
  number={4},
  pages={042132},
  year={2016},
  publisher={APS}
}

@article{demirplak2008consistency,
  title={On the consistency, extremal, and global properties of counterdiabatic fields},
  author={Demirplak, Mustafa and Rice, Stuart A},
  journal={The Journal of chemical physics},
  volume={129},
  number={15},
  year={2008},
  publisher={AIP Publishing}
}

@article{vcepaite2024counterdiabatic,
  title={Counterdiabatic, Better, Faster, Stronger: Optimal control for approximate counterdiabatic driving},
  author={{\v{C}}epait{\.e}, Ieva},
  journal={arXiv preprint arXiv:2403.20267},
  year={2024}
}

@article{hartmann2019rapid,
  title={Rapid counter-diabatic sweeps in lattice gauge adiabatic quantum computing},
  author={Hartmann, Andreas and Lechner, Wolfgang},
  journal={New Journal of Physics},
  volume={21},
  number={4},
  pages={043025},
  year={2019},
  publisher={IOP Publishing}
}

@article{hartmann2020multi,
  title={Multi-spin counter-diabatic driving in many-body quantum Otto refrigerators},
  author={Hartmann, Andreas and Mukherjee, Victor and Mbeng, Glen Bigan and Niedenzu, Wolfgang and Lechner, Wolfgang},
  journal={Quantum},
  volume={4},
  pages={377},
  year={2020},
  publisher={Verein zur F{\"o}rderung des Open Access Publizierens in den Quantenwissenschaften}
}

@article{huerta2025quantum,
  title={Quantum coherence and counterdiabatic quantum computing},
  author={Huerta-Ruiz, Josue R and Araya-Gaete, Maximiliano and Tancara, Diego and Solano, Enrique and Barraza, Nancy and Albarr{\'a}n-Arriagada, Francisco},
  journal={New Journal of Physics},
  volume={27},
  number={8},
  pages={084504},
  year={2025},
  publisher={IOP Publishing}
}

@article{vithanage2025fast,
  title={Fast Transport of Trapped Ultracold Atoms Using Shortcuts-to-Adiabaticity by Counterdiabatic Driving},
  author={Vithanage, Denuwan and Wright, Skyler and Luveina-Joseph, Edith and Larson, Christopher and Samson, Edward Carlo},
  journal={arXiv preprint arXiv:2511.04061},
  year={2025}
}

@article{jarzynski2013generating,
  title={Generating shortcuts to adiabaticity in quantum and classical dynamics},
  author={Jarzynski, Christopher},
  journal={Physical Review A—Atomic, Molecular, and Optical Physics},
  volume={88},
  number={4},
  pages={040101},
  year={2013},
  publisher={APS}
}

@article{takahashi2013transitionless,
  title={Transitionless quantum driving for spin systems},
  author={Takahashi, Kazutaka},
  journal={Physical Review E—Statistical, Nonlinear, and Soft Matter Physics},
  volume={87},
  number={6},
  pages={062117},
  year={2013},
  publisher={APS}
}

@book{liesen2013krylov,
  title={Krylov subspace methods: principles and analysis},
  author={Liesen, J{\"o}rg and Strakos, Zdenek},
  year={2013},
  publisher={Numerical Mathematics and Scie}
}

@article{nandy2025quantum,
  title={Quantum dynamics in Krylov space: Methods and applications},
  author={Nandy, Pratik and Matsoukas-Roubeas, Apollonas S and Mart{\'\i}nez-Azcona, Pablo and Dymarsky, Anatoly and del Campo, Adolfo},
  journal={Physics Reports},
  volume={1125},
  pages={1--82},
  year={2025},
  publisher={Elsevier}
}

@article{kim2022rydberg,
  title={Rydberg quantum wires for maximum independent set problems},
  author={Kim, Minhyuk and Kim, Kangheun and Hwang, Jaeyong and Moon, Eun-Gook and Ahn, Jaewook},
  journal={Nature Physics},
  volume={18},
  number={7},
  pages={755--759},
  year={2022},
  publisher={Nature Publishing Group UK London}
}

\end{document}